\documentclass[a4,11pt,fleqn]{article}
\usepackage[]{graphicx}
\usepackage[margin=1in]{geometry}
\usepackage{hyperref}
\newcommand{\beq}{\begin{equation}} 
\newcommand{\eeq}{\end{equation}} 
\newcommand{\beqa}{\begin{eqnarray}} 
\newcommand{\eeqa}{\end{eqnarray}} 
 
   \def\esim{\mathrel{\rlap{\raise2pt\hbox{$\sim$}} 
    \lower1pt\hbox{$-$}}}         
\def\lsim{\mathrel{\rlap{\lower4pt\hbox{\hskip1pt$\sim$}} 
    \raise1pt\hbox{$<$}}}         
\def\gsim{\mathrel{\rlap{\lower4pt\hbox{\hskip1pt$\sim$}} 
    \raise1pt\hbox{$>$}}}         

\begin{document}

\begin{center}
\Large{\bf Dark 
Matter Evidence, Particle Physics Candidates and Detection Methods}
\end{center}

\author{Lars Bergstr\"om\footnote{E-mail:~\textsf{lbe@fysik.su.se} }}
       
\begin{center}
The Oskar Klein Centre, Department of Physics, Stockholm University\\ AlbaNova, SE-10691 Stockholm, Sweden
\end{center}

\begin{abstract}
The problem of the dark matter in the universe is reviewed. A short history of the subject is given, and several of the most obvious particle candidates for dark matter are identified.
Particular focus is given to weakly interacting, massive particles (WIMPs) of which the lightest supersymmetric particle is an interesting special case and a usful template.  The three detection methods: in particle accelerators, by direct detections of scattering in terrestrial detectors, and indirect detection of products from dark matter particle annihilation in the galactic halo, are discussed and their complementarity is explained. Direct detection experiments have revealed some possible indications of a dark matter signal, but the situation is quite confusing at the moment. Very recently, also indirect detection has entered a sensitivity region where some particle candidates could be detectable. Indeed, also here there are some (presently non-conclusive) indications of 
possible dark matter signals, like an interesting  structure at 130 GeV $\gamma$-ray energy found in publicly available data from 
the Fermi-LAT space detector.  The future of the field will depend on whether
WIMPs are indeed the dark matter, something that may realistically be
probed in the next few years. If this exciting scenario turns out to be true, we can expect a host of other, complementary
experiments in the coming decade. If it is not true, the time scale and methods for detection will be much more uncertain.  
\end{abstract}

\section{Introduction to the Dark Matter problem}

The problem of the dark matter (DM) in the universe is, along with the explanation of ``dark energy'', one of the major still unsolved problems of contemporary cosmology, astrophysics and particle physics. There have been several
extensive reviews over the last 15 years \cite{jkg,lbereview,bertonesilk}. Here
we will only review some of the basic fundaments of particle dark matter, and also present some recent developments, including possible indications of the existence of dark matter experimental signals, which for obvious reasons were not covered in these older reviews.  

Historically, the dark matter problem was for a long time termed the ``missing matter'' problem, and was noted, but not particularly intensely studied, by astronomers from the 1930's and onwards.
Nowadays, the problem is best approached using basic particle physics and our knowledge of the cosmic microwave background, as given, e.g., by the latest results
from the 
WMAP satellite \cite{wmap10}. Soon also the Planck satellite, which according to plans will present data on cosmology in the beginning of 2013, may add 
further details to the picture. 

The particle physics zoo in the early universe was at 
temperatures of the order of a TeV consisting of a plasma of particles in the
Standard Model of particle physics -- quarks, 
leptons and gauge particles (the photon, the electroweak gauge bosons, 
the Higgs particles and gluons). In this plasma, reactions were very 
rapid compared with the expansion rate of the universe, and local 
thermal equilibrium together with adiabatic expansion was for most of the evolution a good 
approximation. There could have been particular epochs when the universe 
underwent phase transitions (see, e.g., \cite{lbe_goobar} for the basic setting in early universe
cosmology. This, however, in general has little importance for the dark matter 
problem. However, for luminous matter, we know from observations that
a very peculiar event has to have happened: the generation of a particle -- antiparticle asymmetry.
The details  here
 are basically unknown, but we know that certain processes -- 
plausibly at much earlier times -- created more baryons than antibaryons, 
i.e., a CP asymmetry occurred, without which the universe as we know it could 
not exist. Without this asymmetry matter and antimatter would have 
annihilated with a large cross section governed by the strong interaction. 
The end result would be basically only nearly thermal photons, 
and the resulting matter and antimatter density would be so 
small that structure would never have formed.

Indeed and fortunately, antimatter does not seem to be present in large quantities in the universe today, 
as can be inferred from the absence of $\gamma$-ray radiation that would be 
created in large amounts if astrophysical anti-objects would annihilate on their matter 
counterparts (this would also cause deviations from the pure black-body form of the
 cosmic microwave background). In fact, both the analysis of primordial nucleosynthesis 
and the CMB give a non-zero number around $10^{-10}$ for the relative baryon-antibaryon asymmetry, 
which means as mentioned  that matter dominated slightly over
 antimatter already in the very early universe. This is confirmed by  the large number 
density of CMB photons, as most of the matter annihilated on antimatter, creating entropy in the form of 
photons.

Thus, thanks to the difference between matter and antimatter, the  
CP asymmetry, 
today in the form of a relic density of baryons, still small, corresponding 
to less than 5 \% of the measured energy density in the universe today, was 
formed, and a  number density of electrons equal to that of 
protons  kept the universe electrically neutral on average. 
The exact mechanism behind the baryon asymmetry is not known. In some current models related to dark matter
a lepton asymmetry was generated first and then 
transferred to the baryon sector 
through non-perturbative processes \cite{leptogenesis}.

The reason that protons, 
electrons and neutrons (hidden in nuclei after big bang nucleosynthesis, foremost in $^4$He) remained, 
and no other known massive elementary particles 
of the Standard Model of particle physics, is related to the short 
lifetime of all other particles. For instance, the muon and the $\tau$ leptons -- the 
"heavier electrons" -- decayed within a microsecond to electrons, positrons, photons and 
neutrinos. (For the $\tau$ the decays after a picosecond  mostly took place through a decay chain containing also $\pi$ mesons. The end
result is the same, as also pions have a very short lifetime.) 
The electron as far as we know is stable, with the lifetime lower limit of more than
$10^{26}$ years \cite{pdg}. 

Here it is appropriate to ask oneself: Why is
 the electron stable? The answer is simple: There is no other particle
 it can decay to, since there is no lighter particle
in nature which has one unit of negative charge. Thus, the stability 
of the electron is guaranteed by two fundamental principles of physics, to which we have found
no exceptions: energy conservation (forbidding a light
 particle decaying to a heavier one) and
the conservation of electric charge. The latter fact is related 
to a powerful unbroken symmetry, the $U(1)$ gauge invariance of electromagnetism to be precise.
For the baryons, such as the proton, there is no deep reason within the Standard Model of particle physics
for stability. There is no gauge symmetry related to baryon number, and there are both lighter positively charged hadrons and leptons that the lightest
baryon, the  proton, could in principle decay into. In fact, the search for proton decay continues,
using the big underground detectors constructed mainly for that purpose, but whose main use, in the absence of detection of proton decays, has been
to study interesting processes in the neutrino sector, such as oscillation between the three neutrino
species in nature. In some 
models of the unification of the gauge forces, proton decay is actually predicted to occur
with a lifetime which is interestingly close to the lower limit of the lifetime of $\sim 10^{34}$ years given by present 
experimental bounds \cite{pdg}.   

There is an important  lesson to learn here: If there exists a family of 
particles in nature which, due to some symmetry, carry a conserved 
quantum number, the lightest of these particles will be stable for 
the seemingly trivial reason that it has no lighter particle with the 
same quantum number to decay to. This is the key to understanding 
how dark matter can have been created in the early universe and filling the universe still today, all 
that is needed is the existence of a "charge". This is not the electric 
charge since electrically charged particles radiate photons and 
thus would be visible, not dark. But there are other 
types of symmetry giving additively or multiplicatively conserved 
quantum numbers.

 The simplest and most important example of such a symmetry which, if unbroken, gives an
infinite lifetime of a particle species is a $Z_2$ symmetry.  $Z_2$ is the
 discrete multiplicative group of
two elements, $+1$ and $-1$. An interesting situation occurs if all ordinary particles of the Standard Model are assigned $Z_2$ 
"charge", or parity, of $+1$.
If the hypothetical family of the ``dark sector'' have parity $-1$,
there has to be one member of the $Z_2$ parity $-1$ 
family that 
is stable, namely the lightest one. This happens for the same reason that the electron is stable: energy
conservation means it can only decay to lighter particles, but if it is the lightest with $Z_2$ parity $-1$ there is no
other state with negative parity it can decay into. This is in fact
the way  a natural dark matter candidate -- the lightest supersymmetric particle -- can appear in theories of 
supersymmetry, the symmetry between fermions and bosons. In many realistic 
supersymmetric models there is indeed
a multiplicatively conserved $Z_2$ quantum number, $R$-parity, which is $+1$ for 
all ordinary particles of the Standard Model and $-1$ for supersymmetric 
partners. Actually, it can be defined by
\beq
R=(-1)^{3(B-L)+2s},
\eeq
with $B$ the baryon number, $L$ the lepton number and $s$ the spin of the particle.
In supersymmetric models, $R$-parity conservation is natural to impose as it
also forbids otherwise potentially catastrophically large baryon number violating 
couplings to occur. 

Of course, we do not 
know if supersymmetry exists (so far there is no hint of its existence
 at LHC -- although this is of course yet at an early stage of LHC operation, still not at design energy). 
However, this mechanism of creating good dark matter candidates 
by having a $Z_2$-like parity symmetry is very general and is one of the key 
ingredients of many of the popular so-called WIMP
models (weakly interacting massive particles) for dark matter. This is one reason why
the lightest supersymmetric particle has attracted much attention as a candidate for 
a dark matter particle - it is an excellent WIMP template, with properties that are 
completely computable (e.g., using the {\sc DarkSUSY} package \cite{ds}) once the (many)
parameters of the model has been specified.

Also in models with universal extra dimensions, Kaluza-Klein (KK) models \cite{profumo},
there appears a ``KK-parity'' which has the value of $+1$ for Standard Model particles
and $-1$ for the first excited states of all Standard Model fields. Then the lightest of the $KK$
parity $-1$ particles has to be stable.

\section{The WIMP scenario}
The reasoning behind the WIMP scenario is as follows (we use here the $Z_2$ symmetry as the simplest one which gives a stable particle, it could of course be a higher discrete symmetry): At high temperatures, i.e., high 
thermal energies in the early universe, particles also with $Z_2$ parity $-1$ 
could rapidly be created or destroyed in pairs,
since a pair has parity $+1$ and can therefore interact with 
ordinary matter, for instance through the reaction
 $\chi\chi\leftrightarrow f\bar f$, with $f$ an ordinary fermion.
Since the dark matter particle $\chi$ has to be electrically neutral,
 we can assume that it is equal to its antiparticle. If it is not, we 
should keep track of antiparticles by also putting a bar over $\chi$ 
in that case. Still, the initial state has $Z_2$ parity $+1$ being the 
product of two negative unit numbers, as has the final state being the
 product of two positive unit numbers. 
It is relatively easy to use early universe thermodynamics to track 
the number density distribution of $\chi$ particles. In the earliest
universe, temperatures were much higher than the particle mass
$T>> m_\chi$ (in units where $k_B=c=\hbar=1$), and the number density of particles
followed a $T^4$ distribution (as in the Stefan-Boltzmann law). When the temperature 
decreased further, the thermal distribution at some point changed from the $n_\chi\sim T^4$ distribution
 to the exponentially suppressed $n_\chi\sim T^{3/2}\exp(-m\chi/T)$ behaviour when
the particles became  non-relativistic, and the particle energy was completely dominated by the rest mass of the particle. Due to the rapidly decreasing number density with decreasing temperature the dark matter particles at some point became too diluted 
to annihilate on each other, and thermal energies were not enough to
pair-create them. 
They therefore left the thermal heat bath with a comoving density 
that did not 
change appreciably after
this ``freeze-out" epoch. 

Analyses combining high-redshift supernova luminosity distances,
microwave background fluctuations (from the satellite WMAP) and baryon acoustic oscillations (BAO) in the galaxy distribution \cite{wmap10,ramme}
give tight constraints on the present mass density of matter in
the Universe. This is usually expressed in the ratio  
$$\Omega_M=\rho_M/\rho_{\rm crit},$$
 normalized to the critical
density, 

$$\rho_{\rm crit}=3H_0^2/(8\pi G_N)=h^2\times 1.9\cdot 10^{-29} \ {\rm
g\, cm}^{-3}.$$

The value of the relic density obtained  from the 7-year WMAP data
\cite{wmap10} for
dark matter consisting of the unknown dark matter particle $\chi$ is 
$\Omega_\chi^2 = 0.113\pm 0.004$, which is around five times higher  than 
the value obtained for baryons, $\Omega_Bh^2 = 0.0226\pm 0.0005$.  
Here $h=0.704\pm 0.014$ is the derived \cite{wmap10} present value of 
the Hubble constant in units of 
$100$~km~s$^{-1}$~Mpc$^{-1}$. As is known, it is 
essentially impossible to fit the WMAP data without the presence of 
cold dark matter, i.e, massive electrically neutral particles 
(particles that were non-relativistic at freeze-out).
In addition, the WMAP data is consistent within one percent with a flat 
universe ($\Omega_{\mathrm tot}=1$) and a value for the dark energy 
component, e.g. the cosmological constant $\Lambda$, of
 $\Omega_\Lambda = 0.73\pm 0.02$.

For WIMPs, the Boltzmann equation 
 can be found by solving the non-linear equation (the Ricatti equation) which 
keeps track of the number density evolution numerically. 
One finds 
that if the WIMPs interacted with regular electroweak 
gauge interaction strength with
 Standard Model particles in the primordial soup, they would have followed the thermal 
distribution until roughly $T_f\sim m_\chi/20$, i.e., they were already quite 
non-relativistic
at this ``freeze-out'' epoch. 
Moreover, to a good approximation, for particles annihilating through 
the $S$ wave the relic density only depends on the total annihilation cross 
section $\sigma_A$ times velocity \cite{jkg},
\beq
\Omega_\chi h^2\simeq 0.11\times
\frac{2.8\cdot 10^{-26}\ {\rm cm}^3{\rm s}^{-1}}
{\langle\sigma_A|{\mathbf v}|\rangle},\label{eq:wimp}
\eeq
where an average of velocities and angles is taken (the correct expression
$\langle\sigma_A|{\mathbf v}|\rangle$ is often abbreviated as 
just $\sigma v$).

The value in Eq.~(\ref{eq:wimp}) for $\sigma v$ evaluates to a cross section 
of 1 pb, i.e., a typical weak interaction cross section.
In other word, if one gives ordinary gauge gauge couplings to the particle $\chi$, and a 
mass of typical weak interaction magnitude (to within an order of magnitude 
 250 GeV, say), then $\sigma v$ is such that the resulting relic
 density comes close to the measured value of $\Omega_\chi h^2\sim 0.11$. 
This is the essence of the WIMP scenario, which is sometimes called the 
``WIMP miracle''. Of course, this may be nothing else than a coincidence, but 
having very few other leads to follow, the WIMP scenario is by far the most followed one, in particular 
as processes of  weak interaction strength that give the correct relic density, in many cases are tantalizingly 
close to cause detection in the rapidly improving current dark matter experiments.

The particle physics 
connection thus is particularly striking in the WIMP scenario. Namely,
 for typical gauge couplings to ordinary Standard Model particles
and a dark matter particle mass at the ordinary weak interaction 
scale, the relic 
density computed using standard early universe thermodynamics (as tested, 
e.g., by the successful calculation of the abundances of light elements
using big bang nucleosynthesis) 
turns out to be in the same range as the cosmologically measured one.
Of course, details may vary somewhat, meaning that WIMP scenarios can be found
with masses between, say, 10 GeV and a few TeV, and in some cases even outside this range. There are deviations possible for the WIMP thermal freeze-out
scenario, for instance there could be other contributions to the 
energy density at a given epoch which would influence the balance between
annihilation rate and the expansion rate in the universe. Also, some of the 
heavier particles in the dark matter sector may influence the relic density. This is particularly important if the mass difference is small, when 
so-called coannihilations may occur \cite{coannihilation}.

Although there is no completely convincing argument for WIMP dark matter -- 
it may as said be a coincidence -- it nevertheless gives WIMP candidates a 
flavour of naturalness. For non-WIMP candidates there is, on the other hand, 
usually a fine-tuning involved, or the use of non-standard early-universe cosmology, to obtain 
the correct relic density. Even limiting oneself to WIMP models for dark 
matter, the literature is extensive \cite{bertone_book}.

Historically, the pioneer of dark matter research was arguably Fritz Zwicky, 
who already in the 1930's pointed out that galaxy clusters, in particular 
the Coma cluster, seem to have too large velocity dispersion among the 
galaxies in the cluster \cite{zwicky} to be held together by their mutual gravity. 
This, using the virial theorem relating kinetic and potential energy, 
would mean that a much larger gravitating mass than the visible mass is
 present, and Zwicky actually coined the term ``dark matter'' for the missing matter, for which perhaps ``invisible matter'' would have been the most correct term. 
We will return to the potential of galaxy clusters for 
pinning down the nature of the dark matter.

We find from the WMAP data the in the universe, that  only 4-5 \% of the energy density is provided by ordinary 
matter, while the remaining 95 \% is composed of two agents we still know 
very little about. Perhaps the biggest mystery is given by the  gravitationally repulsive dark energy (72-73 \%),
for which there is little understanding both of its composition and extremely small size
(by particle physics estimates, which naively would give a magnitude many, many orders of magnitude larger). 

The 
gravitationally attractive dark matter contributes some (22-23 \%) of the total
energy density today.  Whereas additional 
knowledge about dark energy, whether is it a cosmological constant or
 a time-dependent expectation value of a scalar field, or something
 even more exotic, may take more than a decade to gather, dark matter
 may in fact be closer to be identified. 
It can be argued that within a decade we will have tested many of the
 most natural scenarios where the dark matter particles are 
thermodynamically produced in the early universe, in particular the 
WIMP scenario \cite{gianfranco}. This is based on the hope of the development 
on the experimental side to continue at the same impressive pace as in the 
previous decade. Here new experiments, described by other authors in this 
issue, will open new possibilities.  Of course, as the nature of the dark
 matter particle is presently unknown, success can not be guaranteed. The 
advantage of dark matter detectors is, however, for indirect detection 
that it will also study ``bread-and-butter'' physics, from the particle physics point of view, some of which is extremely interesting in 
its own right for the cosmic ray community of physicists and 
astrophysicists. And the development of direct detection experiments
 gives us a whole range of new sensitive detectors of promising potential for 
many fields of high-precision science, such as the search for neutrinoless beta-decays and other potentially revolutionizing
discoveries.

The aim of dark matter research is to explain the measured density of 
dark matter, and to give candidates for the identity of the dark matter 
particles. The fact that dark matter is definitely needed on the largest 
scales (probed by WMAP), on galaxy cluster scales (as pointed out by 
Zwicky, and verified by gravitational lensing and the temperature 
distribution of X-ray emitting gas) all the way down to the smallest 
dwarf galaxies, means that solutions based on changing the laws of 
gravity seem little warranted. In particular, the convincing empirical evidence 
for the existence of dark matter given by the "Bullet Cluster"
 \cite{Clowe:2006eq} is very difficult to circumvent by that kind 
of theories, as the X-ray signal from the baryonic matter and the 
gravitational lensing signal from dark matter are clearly separated 
in these colliding clusters. 

 In fact, despite some recent claims to the contrary, in the standard picture of cold dark matter with standard parameters it is possible to
explain even local Galactic  dynamics near the disk within a few kiloparsecs of the solar system
\cite{tremaine}.
Also, as we have seen, by approaching the problem
from basic particle physics and early-universe thermodynamics, 
the existence of yet to be discovered electrically neutral particles does not
seem very far-fetched. The final and still not given proof of this scenario would of course be  the
 discovery of a dark matter particle induced signal in one or more of the
impressive experiments presently being used or planned for dark matter detection.

\section{Methods of Dark Matter Detection}
There are basically three different methods we may employ to detect dark matter particles , in particular WIMPs, and to gain insight on their mass and interaction strength.
At the moment (in early 2012), the LHC accelerator is gathering data at a very impressive pace, now at 8 TeV in the centre of mass. After a couple of years, the energy will be in the full 14 TeV range that was designed (but which could not be obtained at start-up, due to a technical mishap).  There is great hope that the main experiments at the LHC, ATLAS \cite{atlas} and CMS \cite{cms}, 
may discover events that do not
fit into the current Standard Model of particle physics. It may on one hand be very difficult to prove that one has, for instance, found the dark matter
particles, as the required lifetime has to be demonstrated to be much longer than the age of the universe. In particle detectors, this is of course not possible to prove. However, by getting the general mass scale for events beyond the Standard Model, important input would be given to experiments of the other two types, namely, direct and indirect detection of dark matter. 

Direct detection relies on the fact that dark matter particles should be omnipresent in the 
halo of dark matter in which the Milky Way is embedded, and in particular should travel through the solar system and traverse the earth with typical galactic velocities of $\sim 200$ km/s, i.e., with $v/c\sim 10^{-3}$. From modeling of the Milky Way rotation curve one finds that locally the mass density 
of dark matter should be of the order of $0.4$ GeV/cm$^{3}$, within roughly a factor of two. Combining these numbers one can estimate that, if the cross section is that of a supersymmetric WIMP, scattering on nuclei should take place with a cross section
at or below below a few times $10^{-8}$ pb \cite{ds} (see \cite{goodman} for an early paper on the general idea behind direct detection). In deep underground laboratories, there are a number of ingenious experiments taking data or being deployed with Xenon-100 giving the 
best current bounds, as reviewed by L. Baudis in this issue \cite{baudis_here}.

Direct detection experiments have seen an impressive gain of sensitivity during the last few years, mostly in the low-mass range below 100 GeV mass where they nicely complement accelerator searches. The idea is to register rare events giving a combination of scintillation, ionization and nuclear recoil signals in chunks of matter shielded from cosmic rays in underground sites. There are two factors which mean that direct detection experiments run out of sensitivity at high masses, however. First, the rate is proportional to the number flux of dark matter particles $\chi$ from the halo, i.e., to $n_\chi=\rho_\chi/m_\chi$. Secondly, the scattering cross section is highest when the dark matter particle and the nuclear target have roughly the same mass. However, the decrease in sensitivity
for large masses, being linear, is not very slow but it will be more and more difficult to get an exact estimate of the mass the larger the actual mass is.   

In indirect detection, one rather registers products of dark matter annihilation from regions in the surrounding universe with a high dark matter 
density like the galactic centre, dwarf spheroidal galaxies, or 
galaxy clusters.  
Knowing the halo density distribution, for instance by using numerical simulations (see article by C. Frenk and S. White \cite{frenk_here} in this issue)
one can also estimate the probability that dark matter particles encounter each other and annihilate, creating a new source of  Standard Model particles. 
Of particular importance are antiparticles like positrons and antiprotons (which do not appear with large fractions in cosmic rays due to the baryon asymmetry of our universe). 

Also neutrinos could give an interesting signature as dark matter particles could scatter in the interior of the sun or the earth and be gravitationally trapped there, enhancing the number density and thus the annihilation rate. As two particles have to be at the same place to 
annihilate, the annihilation rate actually is quadratic in the number density of dark matter particles. Depending on the timescale for capture, either the capture rate or the annihilation rate could be the dominant factor. In the dark matter annihilations, typically all Standard Model particles are created. However, all strongly or electromagnetically interacting particles would quickly be stopped in the solar or earth interior, whereas neutrinos would easily
escape (in the case of the sun, this is true at least for dark matter particles with mass less than a TeV -- for more massive particles, absorption in the  solar interior becomes important). The discovery of energetic neutrinos from the direction of the sun would be a spectacular verification of this process,
and the mass of the dark matter particle could then be roughly estimated. The current large neutrino telescopes like IceCube \cite{icecube} and ANTARES 
\cite{antares} have dark matter search in their scientific programmes. For WIMPs captured and annihilating in the earth, the limits obtained from neutrino telescopes are not competitive with direct detection. This is due to the fact that the chemical composition of the earth is dominated by spin-0 elements,
i.e., the same type of elements mostly employed in direct detection experiments. However, protons of course have spin-1/2, and as the sun consists  of hydrogen to some 71 percent of the mass, neutrino telescopes can compete with the current not very much developed experiments targeting spin-dependent scattering of dark matter.  

Recently, most studies of indirect detection have been dealing with $\gamma$-rays from dark matter annihilation. This is due to several unique properties of $\gamma$-rays.
First of all, they do not scatter appreciably during their travel through the galaxy, but rather point back at the site where the annihilation took place. Also, absorption can generally be neglected, as the cross section for scattering on electrons and nuclei for the GeV to TeV range is very small. This means that one may use properties of the energy distribution resulting from these processes to separate a signal from astrophysical fore- or backgrounds.
One obvious consequence of the dark matter origin of a photon generated in dark matter annihilation is, for example, that the photon energy is limited by 
the rest mass of an annihilating particle. As the extreme, but currently interesting case, as will be seen later, is the annihilation into two photons
(as first suggested in \cite{lbe_snellman}). The conservation of energy and momentum means, since the particles annihilate essentially at rest ($v/c\sim 10^{-3}$) that the photons will emerge back-to-back each with an energy of the rest mass of the dark matter particle, $E_\gamma=m_\chi$. There will be a smearing of the photon energy due to the motion on the annihilating pair with respect to the local rest frame, giving a relative width of the order of $10^{-3}$ due to the Doppler effect, but this would of 
course be a spectacular signature. There are two fact that make this process difficult to currently observe, however. First, it is generally induced by 
quantum mechanical loop effects (as the electrically neutral initial state does not couple directly to photons as these only couple  to charged 
particles). This means that the annihilation rate into two photons is suppressed, typically by the factor $\alpha^2$ (with $\alpha$ the electromagnetic fine-structure constant) compared to tree-level processes. Secondly, current detectors have an energy resolution not better than 10 \%, which means that the signal will be smeared and may drown in the background. There
are, however, other processes that are only suppressed linearly in $\alpha$ \cite{lbe89,bbe} and which may help discovery due again to a peaking 
of the photon energy near $m_\chi$. Also, since the 2$\gamma$ line is induced by
a quantumm mechanical loop processes, it may be large if there are many virtual
particles contributing (such as, if they exist, particle with high electric 
charge). 
  
An interesting feature of indirect detection is that the expression for the local annihilation rate of a pair of DM particles $\chi$ (here assumed, like in supersymmetry, to be self-charge-conjugate, of relative velocity $v_{rel}$),
\begin{equation}
\Gamma_{ann}\propto n^2_{\chi}\langle\sigma_{ann}(v_{rel}) v_{rel}\rangle
\end{equation}
is the dependence on the square of the number density. As numerical 
simulations have discovered that galactic halos may be quite crowded with 
sub-halos (see, e.g., \cite{springel}), this means that over the
couple of decades that these processes have been discussed their potential 
for dark matter discovery through annihilation to $\gamma$-rays has 
increased. In fact, there is even a possibility that the entirety of dark 
matter in the universe though this clumping in the evolution of structure 
may generate a cosmological signature 
\cite{cosmic}.

Also, the cross 
section may depend in non-trivial ways on the averaged relative velocity. 
In particular, for low velocities the rate may be much higher than at 
high velocity, for models containing an attractive force between the 
annihilating particles. This is in particular true for models with  
so-called Sommerfeld enhancement \cite{sommerfeld,sommerfeld2}, a resonant 
enhancement by in some cases several orders of magnitude. This means that 
dwarf galaxies (dark matter subhalos) may be particularly interesting objects 
to study, as they are completely dark matter dominated with low rate of 
cosmic ray-induced $\gamma$-rays, and their low galaxy mass means a relatively 
low velocity dispersion. This means higher possible rates if Sommerfeld 
enhancement is active. and this may compensate the $1/m_\chi^2$ decrease 
in the cross section that lies in the factor $n_\chi^2=\rho_\chi^2/m_\chi^2$.
 Some models with Sommerfeld enhancement may seem somewhat contrived, as 
one in general needs almost degeneracy between the dark matter particle 
and a $t$-channel exchange particle. However, as was first pointed out in \cite{sommerfeld}, it exists even in the MSSM, for instance for pure higgsinos 
or winos in the TeV range, where the required degeneracy
 with the lightest chargino generically exists. In fact, it was
 pointed out that 
the otherwise loop-suppressed, "smoking gun" process of annihilation into
 TeV $\gamma$-ray lines may in this case be observable.
 This is one of the interesting features that can increase the $\gamma$ line cross section to make a dark matter search with the Fermi satellite \cite{fermi} and, for TeV dark matter canidates, imaging air Cherenkov telescopes, such as MAGIC, HESS  and the planned large array CTA, very exciting.
 
So far, indirect methods have not been as competitive as direct detection, but recently the Fermi collaboration has started to probe the interesting WIMP region by stacking data from several dwarf galaxies \cite{maja} (see also \cite{koushiappas}).
We will later encounter a couple of claimed, but still uncertain, dark matter signals in $\gamma$-rays.

\section{A WIMP Template: The Lightest Supersymmetric Particle}
Supersymmetry, invented already in the 1970's, and obtained as a phenomenological manifestation of  most of the realistic string theories, has been the prime template for a WIMP since the middle of the 1980's \cite{jkg,goldberg,ellis}. For a variety of reasons, the lightest neutralino,
$$\widetilde\chi^0_1=a_1\widetilde B^0+a_2\widetilde W^0+a_3\widetilde H^0_1 +a_4\widetilde H^0_2, $$
is the most natural choice explaining the dark matter  in the $R$-parity conserving minimal supersymmetric standard model (MSSM). Even in the MSSM, however, there are in principle more than a hundred free parameters, meaning that for practical reasons the templates, for instance in early use at the LHC experiments, were drastically simplified versions (like CMSSM or the even more constrained mSUGRA), which do not, in contrast to the full MSSM, correspond very well to recent models of supersymmetry breaking  \cite{seiberg}. It still remains true, however, that versions of supersymmetric dark matter models are very useful templates when it comes, for instance, to comparing various detection methods. Already now, however, some of the most constrained supersymmetric models do feel quite a large tension from LHC data \cite{lhc_nosusy}. 

Even in less simplified versions, like the 19 to 24-parameter "phenomenological MSSM" \cite{pmssm}, as can for instance be treated in the computer package DarkSUSY \cite{ds}, the bounds on particle masses given, e.g., by fulfilling the WMAP relic density, are not very constraining at the moment \cite{allanach}. In particular, dark matter candidate neutralinos can be found in the several TeV range, and of course a general WIMP may be even heavier, perhaps approach the unitarity bound of several hundred TeV \cite{unitarity}, although for such high masses in the MSSM one really would have to worry 
about the subjective issue of fine-tuning of 
masses and couplings -- something that really would go against the spirit of introducing supersymmetry in the first place. Of course, the outlook for the MSSM would be much bleaker if a light Higgs (with mass below roughly 130 GeV) were not to be found by the end of the 8 TeV run at LHC, in 2012. (Although the Higgs sector may be more complicated, and invisible decays may be present, for example.) It is interesting to note, although still to early to judge the significance of, the early indications of a Higgs signal at a mass around 125 GeV \cite{atlas_125,cms_125}. This is indeed within the possibility of the MSSM, although such a large radiative correction to the mass again puts some strain on the naturalness of parameters.

\subsection{Non-WIMP Dark Matter Candidates}

Taking WIMPs  as  the leading candidates for dark matter, inspired by the lack of
fine-tuning to get correct relic density, we  expect an angle- and velocity-averaged cross section times relative velocity of
$$(\sigma v)_{\rm WIMP} \sim 3\cdot 10^{-26}\ {\rm cm}^3{\rm s}^{-1}.$$

Of course a word of caution is in place here. There are many non-WIMP models that also have some particle
physics motivation, and may be detectable, like:  axions, gravitinos, superWIMPS, non-thermal dark matter, decaying dark matter, sterile neutrinos, Q-balls\ldots
For a thorough review of the status of the field see, e.g., the recent 700-page extensive review of these as well as more standard WIMP models \cite{bertone_book}.  For a less ambitious, more recent review with focus on imaging air Cherenkov telescopes, see \cite{lbe12}.

Noticeable progress the last few years has especially been made in axion searches \cite{axions} -- but again, no dark matter signal has been found yet. In the following, we will deal mainly with the main templates - WIMPs.

There may be non-WIMP dark matter giving some contribution to the relic density, like additional sterile light degrees of freedom that would be ``hot'' and according to some analyses even may be preferred in cosmological
 data \cite{hamann}. However, the significance for this is not strong and may
 be influenced by statistical bias \cite{verde}. Another possibility is ``warm'' DM \cite{sterile} consisting of keV-scale sterile neutrinos. There the production rate in the early universe generally has to be tuned if one wants to have 
the observed relic density. Also here, there are some arguments in favour of this scenario \cite{lovell}. 

Ordinary, active 
neutrinos have too small mass to contribute significantly to the dark
 matter density, although in the extreme case may contribute a couple
 of percent to the critical density today. With the upcoming data from 
the Planck satellite we will hopefully gain more knowledge on the cosmological
aspects of the neutrino sector. 

\section{Indirect Detection through $\gamma$-rays}

At the moment, indirect $\gamma$-ray dark matter detection is evolving very rapidly,
in particular thanks to the successful operation of the Fermi-LAT space 
detector, and also the advanced plans for a very large imaging air Cherenkov
telescope array, superseding the successful detectors HESS \cite{hess}, MAGIC \cite{magic} and VERITAS \cite{veritas}, namely the CTA \cite{CTA,lbe12}.

For WIMP models, there have recently also been 
improvements in the estimates of the annihilation rate and 
the $\gamma$-ray flux expected from annihilations in the galactic halo.
In the galaxy, the dark matter particles move extremely slowly by particle physics standards, as $v/c\sim 10^{-3}$. An interesting effect is
 caused due to the suppression of the $^3S_1$ state for a system of two 
Majorana spinors (such as neutralinos) annihilating at slow velocity. This 
is due to the requirement of Fermi statistics for the two identical fermions. In this state the two fermions have parallel spins, which becomes forbidden as the velocity tends to zero.    
This means that annihilation only
 occurs from the pseudoscalar $^1S_0$ state where spins are opposite, causing for instance the 
annihilation amplitude into a light fermion-antifermion pair, 
such as $e^+e^-$,  to be suppressed by an explicit helicity  factor of 
the fermion mass. Direct annihilation into $e^+e^-$ was thus thought 
to be very subdominant \cite{goldberg}. However, it was realized \cite{bbe} (building
 on an old idea \cite{lbe89}), that a spin-flip by one of the Majorana 
fermions caused by emitting a photon (so-called internal bremsstrahlung) 
could first of all allow evasion of the 
helicity suppression of the  process to a mere $\alpha/\pi$ ordinary 
radiative factor.  And, in addition, the spectral shape of the emitted 
photon is very favourable for detection, increasing rapidly with photon
energy causing a peak
 close to the dark matter particle mass. In particular, for heavy 
(TeV-scale) WIMPs this could be quite important, and using the 
radiative peak would help extracting the signal over background 
\cite{desy}. Recently, these radiative processes have been generalized 
also to emission of other gauge bosons, and have been shown to be quite 
important generally \cite{radiative}. 

One difficulty when estimating $\gamma$-ray rates from dark matter 
annihilation is the still poorly known distribution of halo dark matter on 
the smallest galactic and subgalactic scales. 
$N$-body simulations indicate, but with limited space resolution, that 
the halo should be very abundant with dark matter clumps \cite{springel}. Due to tidal disruption of these subhalos close to the galactic centre, the distribution is particularly important in the outer parts of galactic halos. The general 
expression for the annihilation signal is a line-of sight integral along a given direction which 
is proportional to the square of the local density along the way \cite{bub},
 \begin{eqnarray}
{\Phi_{\gamma}(\psi)\over \rm{cm}^{-2}\;\rm{s}^{-1}\;\rm{sr}^{-1}} & \simeq & 0.94 \cdot 10^{-13}
\left( \frac{<\sigma v>_\gamma}
{10^{-29}\ {\rm cm}^3 {\rm s}^{-1}}\right)\left( \frac{100\,\rm{GeV}}
{M_\chi}\right)^2 
J\left(\psi\right) \nonumber \\
\end{eqnarray}
with the astrophysical part residing in  the dimensionless function 
\begin{equation}
J\left(\psi\right) = \frac{1} {8.5\, \rm{kpc}} 
\cdot \left(\frac{1}{0.3\,{\rm GeV}/{\rm cm}^3}\right)^2
\int_{l.o.s.}\rho^2(l)\; d\,l(\psi),
\label{eq:jpsi}
\end{equation}
where enhancements of the density even on very small scales may be important. 
An example, besides the subhalos residing in the outer parts of the halo, is the region near the galactic centre.  There the gravitational field is 
dominated by the black hole and a stellar cusp is known to exist, 
with unknown effects 
on the dark matter density and therefore the 
annihilation rate into $\gamma$-rays. Unfortunately, the contribution 
from dark matter to the  rotation curve is much too small in the inner parts 
of the galaxy to enable to determine whether the halo density is cuspy or not. 
Also, even a very important concentration of dark matter very near the black hole could well influence the indirect detection rate \cite{gondolo_silk} without
having much of a dynamical effect otherwise.  

A cuspy behaviour would be
favoured by the results of $N$-body dark matter-only simulations, but  
it is possible that it has 
a milder dependence or even a core \cite{sofue}. As a simple template of 
the DM distribution, an NFW profile \cite{nfw} 
having an $1/r$ cusp near the center is often 
used (or a very similar, 
so-called Einasto profile \cite{einasto}). 
This has recently been shown to be consistent with data on 
microlensing in central regions of the galaxy \cite{iocco}. The uncertainty
 very near the galactic centre remains, however, and it is unknown whether
 the real distribution is more cuspy or less, and predictions thus vary by
 several orders of magnitude. Even if the center of the galaxy plausibly
 is the most interesting place to search for $\gamma$ rays from DM 
annihilation, fore- and backgrounds from astrophysical processes may 
be large, and thus it may be advantageous to search in directions 
close to, but not exactly at, the galactic center \cite{preglast}. 

Interesting objects are the dwarf  galaxies mentioned above, where  limits
from Fermi-LAT 
are now getting into the parameter space of common WIMPs \cite{maja,koushiappas}. The 
abundance of DM  in these dwarf galaxies is much higher, as star formation probably
 only occurs above some threshold mass, and the pressure from a few supernovae
 may be enough to completely empty a dwarf galaxy from baryons. This means that there may  be  bright spots in $\gamma$-rays in the sky, ``dark matter clumps'', that are not visible at all in any wavelength, except their possible annihilation to $\gamma$-rays. 

Simulations indicate that 
DM clumps will be destroyed by tidal forces near the center of galaxies but
 can be very abundant in the outer regions \cite{springel}. 
When going to larger scale objects like galaxy clusters that are forming at the present epoch, the number 
of undestroyed DM clumps may be even larger, making these clusters -- in 
a perhaps unexpected agreement with the discovery of Zwicky -- quite 
promising targets for indirect searches \cite{clusters}.

In the first runs at LHC, no signs of a Higgs particle, nor supersymmetry 
or any other of the prime candidates for dark matter, have yet been discovered.
 On the other hand, the mass region 115 - 130 GeV, interesting for the
 lightest Higgs boson in the simplest versions of supersymmetry, still has to 
be thoroughly investigated. One possible scenario might be that such a Higgs particle 
will  indeed be found, but the particles carrying non-trivial $R$-parity  all 
have masses beyond reach with the LHC. This is not an impossible scenario, 
however it depends 
on the amount of fine-tuning one is willing to tolerate. If one
 puts no prior constraints on the supersymmetric parameter space other than 
one should have the WMAP-measured relic density, and fulfill all other 
experimental constraints ({\em cf.} \cite{abdussalam}), a mass for the 
lightest supersymmetric neutralino in the TeV region is generic. For such
 heavy dark matter neutralinos, the rate for direct detection will also be 
small, and it would seem very difficult to  test such a scenario in 
the foreseeable future. However, for this particular case indirect 
detection through $\gamma$ rays turns out to have an interesting 
advantage, as  imaging air Cherenkov arrays like CTA \cite{CTA} will
have their peak sensitivity in the energy range between a few hundred
 GeV to a few TeV. Depending on the particular dark matter model realized in nature, 
Sommerfeld enhancement of indirect detection may also be operative. For recently
obtained 
realistic estimates of the reach of CTA, see \cite{cta-new}.

A large array like CTA will be served by a diverse astrophysical community with 
interest, e.g., in transient or periodic events, meaning 
that the ``boring'' search for a stationary dark matter spectral signature 
during hundreds or even thousands of hours seems difficult to imagine. One
 may therefore consider, for the first time,  a dedicated particle 
physics  experiment with the search for dark matter as its only aim. One such preliminary study proposed the  
``Dark Matter Array'', DMA \cite{dma}, only to be used for dark matter search. 
Such a dedicated instrument for indirect detection could have great, and 
complementary, potential compared to the large direct detection experiments presently planned. 
There are preliminary ideas \cite{aharonian} 
on how to decrease the lower threshold for detection, which 
could increase the sensitivity for DM detection considerably. In the first
 analysis \cite{dma}, it was investigated how a next-generation, particle
 physics type detector, the DMA, would improve limits of, or detect, dark 
matter. A size ten times that of CTA was considered, having a low energy 
threshold of 10 GeV. Moreover, by making dedicated observations of promising
 dark matter 
targets like dwarf galaxies, where the low energy threshold and good energy resolution help to reject various sources of fore- and backgrounds, one 
would be able to increase discovery potential considerably. Also studies 
of the galactic centre and possibly galaxy clusters, with a total of
 perhaps 5000 hours of observation in 5 years would open interesting 
new vistas in indirect detection. 

For the analysis of the Fermi-LAT discovery potential, 
the approach of \cite{preglast} was followed in \cite{dma}.  20 
logarithmic bins were considered in the $1-300\,$GeV energy range. 
The sensitivity, angular resolution, energy resolution and 
effective area were taken according to Fermi-LAT official specifications. 
The observation time of Fermi was set to five years. To compare with the CTA, 
an energy threshold for CTA of 
$40\,$GeV was used, and the same logarithmic bin size as for Fermi was 
assumed, up to the few TeV range. The projected sensitivity 
curves of \cite{CTA} was assumed, with an angular resolution of 0.02 
$\deg$ and, as usual for dark matter targets in present-day Cherenkov 
arrays, an exposure time of 50 h. The effective area is about $3\,$km$^2$ 
at $5\,$TeV, $1\,$km$^2$ at $100\,$GeV and a fall-off for lower energies 
down to $\sim0.1\,$km$^2$ at $40\,$GeV. The low-energy characteristics of
 CTA will depend on detector details, still to be decided. For a more recent, 
realistic evaluation of the reach of CTA, see \cite{cta-new}. 

For studies of the the galactic center, it was assumed in \cite{dma} that the angular 
resolution of upcoming ACTs will be sufficient to discriminate sources 
like the HESS source very close to, but off-set from, the galactic center 
and which 
is (probably) unrelated to dark matter. For the diffuse background, the
 model developed by the Fermi-LAT group that describes the so far available
 data was used \cite{digel}. 
It was required that the significance in the best energy bin is at least at
 the $5\sigma$ level, i.e.~$S/\sqrt{S+B}>5$, in order to claim that a 
$\gamma$-ray experiment is able to rule out a given dark matter model.
A crude estimate of the potential of a 1t Xenon direct detection
experiment was also made, 
for more details, see
\cite{dma}.

\begin{center}
\begin{figure*}[t!]
  \begin{minipage}[t]{0.9\textwidth}
\begin{center}
  \includegraphics[width=\textwidth] {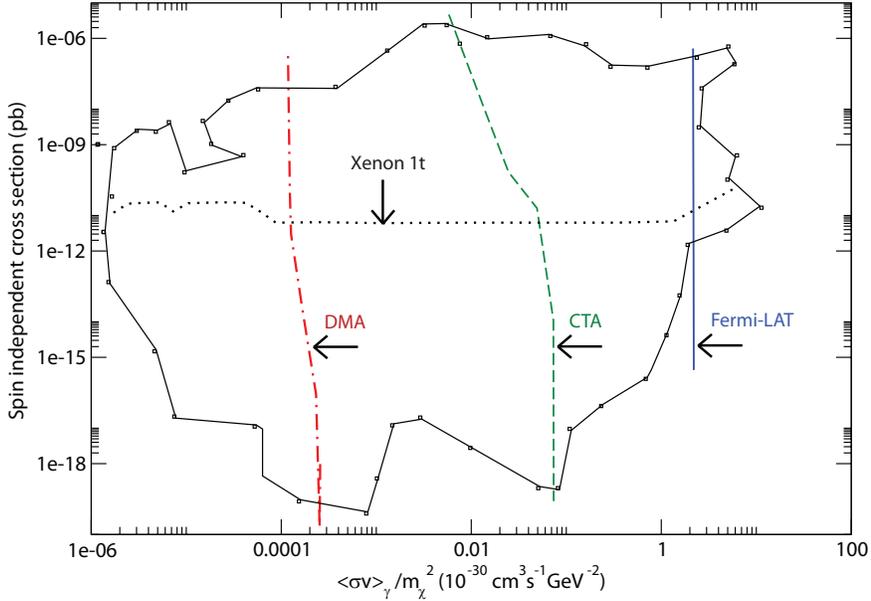}
\end{center}   
\end{minipage}
\caption{Illustration of the reach of direct and indirect dark matter detection experiments. Here $\gamma$-ray detection towards the galactic center with the NFW profile is considered. Shown is the area encompassing the 
approximate range of WMAP-compatible phenomenological MSSM model space, and the reach of the upcoming Xenon 1t direct detection experiment, and the Fermi-LAT, CTA and DMA indirect detection experiments. For details, see  \protect\cite{dma}.}
\label{fig_gc_lbe}
\end{figure*}
\end{center}

We see (Fig.~\ref{fig_gc_lbe}, {\em cf.} \cite{dma}) that a very interesting complementarity could be reached by combining the indirect detection data with direct detection experiments. Actually, imposing CERN LHC bounds at the end of 2011 on squarks and gluinos do not change the region shown in Fig.~\ref{fig_gc_lbe} very much. If a working prototype of the DMA type could be built, this idea may materialize in the next decade as a new way to search for phenomena beyond the Standard Model -- with an expensive dedicated detector, albeit still far below the cost of a new high-energy accelerator. 

\section{Claims of Possible Detections of Dark Matter}
There have recently been a number of claimed possible detections of dark matte (see Table~\ref{tab:tab1} for a summary):
 First of all there is the annual modulation found in the more than decade-old DAMA/LIBRA experiment \protect\cite{dama}. 
The claimed excess of the annual modulation (predicted in \cite{freese} as a promising dark matter detection method) is approaching 9$\sigma$. This is still unexplained at the moment, but is unfortunately not confirmed by other experiments \cite{cdms,xenon100,kims12}. The same is true for the 
 CoGeNT excess events and annual modulation \cite{cogent}, which also is in tension  with the same data, claimed to be more accurate \cite{cdms,xenon100}.  One should be aware, however, that this area of investigation is at present 
beset with large controversies, and one should allow the dust to settle before drawing strong conclusions in 
either direction. At any rate, explanations of these experiments in terms of dark matter scattering seem to probably 
take one out of the standard WIMP scenario or use non-standard dark matter 
halo models (see, e.g., \cite{schwetz, frandsen}, but also \cite{bottino}). The experimental situation is improving rapidly (see \cite{baudis_here}) and it will be very interesting to see if a coherent picture will materialize. An interesting idea has been put forward to use the IceCUBE site at the South Pole to deploy an experiment, DM-Ice, with 
the same type of  NaI crystals as DAMA/LIBRA  uses. A prototype is already 
being tested in the ice \cite{dm-ice}. If the modulation obviously seen in the 
DAMA/LIBRA data is caused by (unknown) environmental, seasonal,  effects, the phase of the modulation should change at the South Pole. On the other hand, the same pattern
as that seen by DAM/LIBRA should reveal itself if this is a genuine dark matter
scattering effect.

\vskip .5cm
\begin{table}
\begin{center}
\begin{tabular}{|p{7cm}|p{7cm}|}
\hline
{\bf  Experiment} & {\bf Status of claim}\\ \hline
\hline DAMA/LIBRA annual modulation\protect\cite{dama} & Unexplained at the moment; not confirmed by other experiments \cite{cdms,xenon100,schwetz,frandsen}\\
\hline CoGeNT excess events and annual modulation \cite{cogent} & Tension with other data \cite{cdms,xenon100,schwetz,frandsen}\\
\hline CRESST excess events \cite{cresst} & Tension with other data \cite{cdms,xenon100,schwetz,frandsen}\\
\hline EGRET excess of GeV photons \protect\cite{egret,wim}& Due to instrument error (?) -- not confirmed by Fermi-LAT \protect\cite{fermi_diffuse}\\
\hline INTEGRAL 511 keV $\gamma$-line from galactic centre region \protect\cite{integral}& Does not seem to have spherical symmetry -- shows an asymmetry which follows the disk (?) \protect\cite{integral_new}\\  
\hline PAMELA: Anomalous ratio of cosmic ray positrons/electrons \protect\cite{pamela}& May be due to DM \cite{pamela_th}, or pulsars \cite{pulsars}
-- energy signature not unique for DM\\
\hline Fermi-LAT positrons + electrons \protect\cite{fermi}& May be due to DM \cite{pamela_th}, or pulsars \cite{pulsars} -- energy signature not unique for DM\\
\hline Fermi-LAT GeV $\gamma$-ray excess towards galactic centre \protect\cite{hooper_goodenough}& Unexplained at the moment -- astrophysical explanations possible \cite{felix,boyarski}, no statement from the Fermi-LAT collaboration\\
\hline WMAP radio ``haze'' \cite{fink}& Has a correspondence in  ``Fermi bubbles'' \cite{su} -- probably caused by outflow from the galactic centre\\   
\hline $\gamma$-ray structure \protect\cite{han} in public Fermi-LAT data \protect\cite{fermi_lat} from galaxy clusters. & Very weak indication, could be cosmic-ray induced emission?\\ 
\hline $\gamma$ line at 130 GeV \protect\cite{bringmann,weniger,tempel} in Fermi-LAT public data \protect\cite{fermi_lat} & $3.3\sigma - 4.6\sigma$ effect, unexplained at the moment. Not confirmed by the Fermi-LAT collaboration \protect\cite{fermi_line_search}.\\
\hline     
\end{tabular}
\caption{Some of the recent experimental claims for possible dark matter detection, and a comment on the present status.}
\end{center}\label{tab:tab1}
\end{table}
\vskip .5cm
   
The recent improvement of the upper limits on the WIMP-nucleon scattering cross section reported by CDMS II \cite{cdms} and, in particular, XENON100 \cite{xenon100} are truly impressive. Not only does it cast some doubt on other reported experimental results, the sensitivity is also good enough to start probing the parameter space of realistic supersymmetric models \cite{ds}. The new calibration of the sensitivity to low-energy recoils of liquid Xenon, although not undisputed \cite{collar}, would seem to
add to the credibility of the new upper limits \cite{xenon100} for the WIMP-nucleon scattering cross section. The very good news is also that the installation of the next stage, a 1 ton liquid Xenon detector, has already started in the  Gran Sasso experimental halls in Italy \cite{xenon1t}.

An early possible indication of a dark matter signal in indirect detection was the EGRET excess of GeV photons \protect\cite{egret,wim}. However, this was not confirmed by the recent much superior data from Fermi, more exactly the large area $\gamma$-ray telescope part of Fermi, Fermi-LAT, and was probably due to instrument error \protect\cite{fermi_diffuse}. 

Another possible indication of a dark matter signal was the discovery of  by INTEGRAL of a 511 keV $\gamma$-line from the galactic centre region \protect\cite{integral}. However, in this energy range positron emission from other sources is possible, and the excess  does not seem to have the spherical symmetry around the galactic centre expected from dark matter annihilation. Rather, it shows an asymmetry which  seems to follow the disk \protect\cite{integral_new}.  Also there are bounds from $\gamma$-ray emission at slightly higher energies that make a DM explanation difficult \cite{yuksel}. It would seem to require a DM candidate in the few MeV region.

A few years ago, a rather strong $\gamma$-ray source, consistent with a point source, was detected from the direction of the galactic centre by the HESS experiment \cite{hess}. Although many dark matter models were tested, it turned out that the shape of the spectrum (and also the inferred mass, which would exceed some 10 - 20 TeV) made quite poor fits to the distribution of dark matter-induced photons.    

Dark matter annihilation occurs from a matter-antimatter symmetric initial state and, barring larger CP-asymmetric couplings, equal amounts of  matter and antimatter should be created. This leads to an interesting possible primary source of stable anti-particles, positrons and antiprotons, in the cosmic rays of dark matter halos, including the  Milky Way halo. (There is in addition a small, 
usually 
negligible, amount of antimatter produced as secondary particles in 
collisions with galactic gas and dust by ordinary cosmic rays.) This possibility of dark matter detection was 
extensively 
discussed a few years ago \cite{strumia,lbnewrev} as the PAMELA and Fermi-LAT
 collaborations had just discovered an anomalously high ratio of positrons 
over electrons up to 100 GeV \cite{pamela}, and sum of positrons and electrons
 up to 1 TeV \cite{fermi}, respectively. During the last years, this anomaly,
 although possible to explain by dark matter annihilation,  was shown to 
need very large boost factors (e.g., from Sommerfeld enhancement), and  
somewhat contrived, 
leptophilic models. On the other hand, certain astrophysical explanations of these observations 
are possible with quite standard assumptions. One cannot say that the dark 
matter explanation is yet ruled out, but it sees tension from other 
measurements, especially from $\gamma$-ray data \cite{fermi_sthlm}. 
It may thus in principle be due to DM \cite{pamela_th}, with rather unusual
 properties, or a more mundane source such as  creation of $e^+e^-$ pairs 
from pulsars other other 
supernova remnants \cite{pulsars}.
It is an unfortunate fact that the energy signature, which in some models with 
large annihilation rate seems to fit the measured energy distribution 
very well, \cite{pamela_th}
is not unique for DM. 

By analyzing public Fermi-LAT data, an excess in the few GeV $\gamma$-ray 
energy region has recently been proposed to be caused by dark matter annihilation. This is extended emission corresponding to 
a $\gamma$-ray source in the region of the galactic centre 
\protect\cite{hooper_goodenough, hooper_linden,hooper_12}. Again, 
this possible dark matter  effect (with a mass around 10 GeV) 
is unexplained at the moment, but 
astrophysical explanations are of course possible in this crowded part
 of the galaxy \cite{felix,boyarski}. It would gain 
significance considerably if the Fermi-LAT collaboration  with their
experience of modeling possible astrophysical sources would confirm this unexplained excess. At the moment there 
is no such statement from the Fermi-LAT team, however.

Even more recently, a $\gamma$-ray excess from galaxy clusters ascribed to dark matter 
has also been claimed 
again using public Fermi data \cite{han}. Although not unexpected
\cite{clusters}, the existence of this (weak) signal relies heavily on the
estimate of the cosmic ray-induced photons, which probably is not 
accurate enough for this purpose. 

Another excess, the WMAP radio ``haze'' found by Finkbeiner and 
collaborators, \cite{fink} has recently been seen to have a correspondence 
in the GeV range in the so-called ``Fermi bubbles'' \cite{su}. These are 
sharply defined regions, several tens of degrees across, on both sides of 
the galactic disk that seem to originate from the galactic center, by 
some so far unknown process. This is again a discovery that has  been 
made using public Fermi data, but it seems that it is difficult to fit 
the shape and the sharp edges of the bubbles with a dark matter 
origin \cite{dobler}. 
The emission is  more plausibly caused by bipolar outflow of cosmic rays 
(electrons or 
protons) from the galactic center, maybe as an episodic event a few million years ago. The existence of the Fermi bubbles adds to the complexity when 
analyzing the $\gamma$-ray flux from the galactic center region, which 
otherwise would be the most natural one where to search for a 
dark matter signal. 
The $\gamma$-ray spectrum seems to be very flat in intensity over the bubble 
regions, with a flux of \cite{su}
\begin{equation}
E_\gamma^2\left(\frac{dN}{E_\gamma}\right)_{\rm bubble}\simeq 0.3\ {\rm keV}\ \rm{cm}^{-2}\rm{s}^{-1}\rm{sr}^{-1}, \label{eq:bubble}
\end{equation}
up to somewhere between 120 and 200 GeV, where the spectrum becomes much softer.

\begin{figure*}[t!]
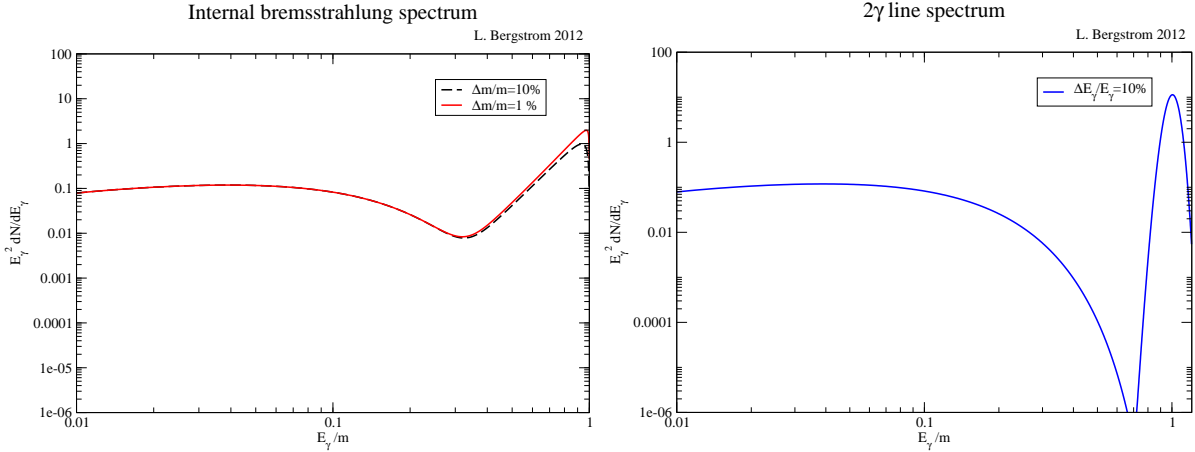

  \begin{minipage}[t]{0.99\textwidth}
\begin{center}
  \includegraphics[width=0.48\textwidth] {weniger.eps}
\includegraphics[width=0.48\textwidth] {annalen.eps}
\end{center}   
\end{minipage}
\caption{Example of internal bremsstrahlung spectrum (left figure) and 
2$\gamma$ line spectrum (right figure). For both processes a typical continuum 
spectrum from quark fragmentation has been included. For the line an
 arbitrary relative normalization of the two has been assumed, and a 
10 \% energy resolution of the detector. As can be seen, both processes
 give a very characteristic rise in the $E^2dN/dE$ distribution near the
 kinematical endpoint $E/m=1$.}
\label{fig:spectra}
\end{figure*}

A problem with using the continuous $\gamma$-ray spectrum for dark matter detection is that for photons coming from
the fragmentation of quarks or gauge bosons, the energy distribution is rather featureless, besides that fact that at
low energies it is harder than $E^{-2}$ and close to the endpoint at $E=m_\chi$ it falls very fast. However, much more spectacular signatures can be found, namely the 2$\gamma$ line \cite{lbe_snellman} (or $Z\gamma$ line \cite{kaplan})  which gives
a spike at   $E=m_\chi$ or $E=m_\chi(1-m_Z^2/(4m_\chi^2))$ respectively, and internal bremsstrahlung \cite{lbe89,bbe} which, for self-conjugate Majorana fermions, also gives 
a steeply rising spectrum near the kinematical endpoint. Examples of these 
rather striking energy distributions are given in
Fig.~\ref{fig:spectra}.

Very recently, an intriguing feature has been found \cite{bringmann,weniger} 
(see also \cite{tempel}), again by using public Fermi-LAT data \cite{fermi_lat}.
Namely, a structure similar to what is expected for internal bremsstrahlung,
 or alternatively a narrow $\gamma$-ray line, is visible in the
 energy range between 130 and 150 GeV. (For the line interpretation the dark matter mass is close to 130 GeV, for internal bremsstrahlung the fit is closer to 
150 GeV. Of course, the spectrum may also be a combination of the two with arbitrary relative strength.)

\begin{figure*}[t!]
\begin{center}
  \begin{minipage}[t]{0.6\textwidth}
  \includegraphics[width=\textwidth] {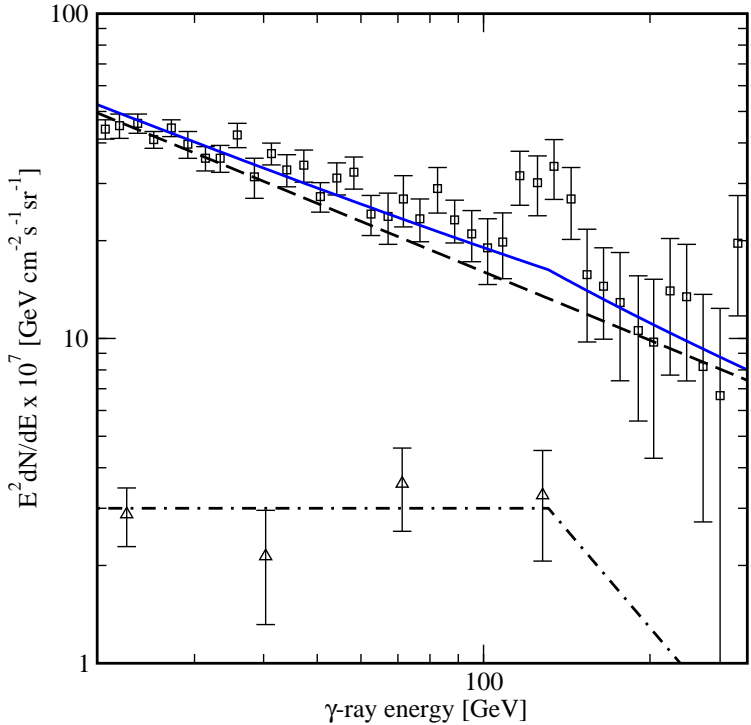}
\end{minipage}
\caption{The public Fermi-LAT \cite{fermi_lat} data extracted from 
\protect\cite{weniger} (squares; ``Reg.4, SOURCE class'', in
\protect\cite{weniger}) as well as the  spectrum of the 
Fermi bubbles \protect\cite{su} (triangles). The dashed line is a
 featureless $E_\gamma^{-2.7}$ spectrum, 
and the dash-dotted line is a simple fit to the Fermi bubble data,
 where the actual location of the break and the slope above the 
break are unknown. The solid line is the summed spectrum (power-law 
plus bubbles), assuming that the break is coincidentally at the same 
energy as the line excess.}

\label{fig:weniger}
\end{center}   
\end{figure*}

These studies used a very interesting 
new method to search for dark matter lines. The region analyzed was allowed
 to vary according to the model used for the dark matter density distribution 
in the galaxy, to maximize the signal to noise ratio. Since the regions such 
identified bear some resemblance to the Fermi bubbles, it was soon conjectured
 that the sudden fall-off of the bubble spectrum in this energy range could
 make one mistake a step in two superimposed continuous distributions for a
 line \cite{profumo12}. However, the bubble flux in Eq.~(\ref{eq:bubble}) is too small to influence 
the estimated line flux appreciably. This is shown in Fig.\ref{fig:weniger},
 where the structure found in \cite{weniger} is displayed and clearly seen, and a good fit 
to the background is given by a continuous power-law falling as $E_\gamma^{-\beta}$
 with $\beta = 2.7$ and where the bubble flux Eq.~(\ref{eq:bubble}) has been added.This conclusion of a good fit
depends very little on the energy where the fall-off of the bubble flux is 
assumed to begin, and also on the steep slope of the bubble energy spectrum
after the fall-off point. 

Although perhaps a tantalizing indication of a dark 
matter signal, this finding of a $\gamma$ line seems to be in slight tension with a recent
 line-search analysis of the Fermi-LAT team
\cite{fermi_line_search}, although as explained in \cite{weniger} this could be due to the different methods used in the analysis. See also \cite{boyarsky_n}
for further comments about alternative explanations. 

With Fermi-LAT taking data for some five more years, we will definitely 
be able to follow this interesting development in detail. 

At energies, say, below 150 GeV, the Fermi-LAT instrument will be 
difficult to surpass by using imaging air Cherenkov telescopes like the CTA, 
which have their greatest sensitivity in the TeV region. However, an 
interesting player
has recently entered the dark matter indirect detection scene. This is 
the Russian-Italian project GAMMA-400
 \cite{gamma-400}, which has a launch around 2018 approved and is planned 
with a slightly smaller effective area than Fermi-LAT, but with better angular 
resolution and in particular 
better energy resolution than that of Fermi-LAT by  an order of magnitude. 
This would take the search for dark matter to another level of sensitivity. 
If the present indication of a line signal would persist, it should be seen 
in GAMMA-400 with a significance of the order of $10\sigma$ \cite{bbcfv}.   
 
If one would assume
that the effect is real and originating from the 2$\gamma$ line one could
speculate  about other measurements that would give more information about
the dark sector. For instance, in the type of models,  e.g., explored in
\cite{baltz}, which connects the dark matter problem with that of neutrino
masses, one would both have a 2$\gamma$ line and an internal bremsstrahlung
feature, which may appear with more statistics. However, there the charged scalar contributing to the emission is an isoscalar, meaning that no corresponding $Z\gamma$ line \cite{kaplan} would
be seen. It would not be difficult to extend the dark sector such that, e.g.,
a scalar isotriplet contributes 
(maybe with charge 2, which would make the line stronger
by a factor of $2^4=16$, as the two gamma vertices in the amplitude would each be a factor of 2 larger, and thus the cross section 16 times higher). In general,
one would then expect also a $Z\gamma$ line (with intrinsic width around a 
percent, due to the finite width of the $Z$) at 
\begin{equation}
E_\gamma(Z\gamma\ {\rm line})=m_\chi\left(1-\frac{m_Z^2}{4m_\chi^2}\right),\label{eq:zgamma}
\end{equation}
which for a mass $m_\chi=130$ GeV corresponds to around $E_\gamma=114$ GeV. It may also be amusing to speculate that the 130 GeV dark matter candidate connects to the 125 GeV Higgs candidate indicated by preliminary LHC data. Then the process $\chi + \chi \to H + \gamma$ may be searched for, with an energy (replacing $m_Z$ by $m_H$ in Eq.~(\ref{eq:zgamma}))  of 100 GeV. If such a signal were to be found, it would show that the dark matter particle $\chi$ would neither be scalar, pseudoscalar  or a Majorana fermion (as these annihilate only from a spin-0 state in
the $S$ wave, and then the $H\gamma$ final state would violate spin conservation). The dark matter particle thus would have to be a Dirac fermion or a spin 1 boson (or a higher spin representation).
  
Of course, model building may seem premature for an effect that may turn 
out to be a statistical fluctuation. The important point to make, however, is that once a dark matter
signature is found, it opens a whole area of further measurements in indirect
detection, as well of course for direct detection. To be finally convinced
that the dark matter problem has indeed been solved, several independent
measurements may be needed, using all three methods we have at our disposal:
accelerator searches, direct and indirect detection, which we have seen all have
a nice complementarity.  

Thus, of all these possibilities of dark matter being detected, the most interesting one at present, besides
the DAMA/LIBRA-CRESST-CoGeNT rather confusing situation (but see 
\cite{frandsen1,kelso,hooper_12}), are the possible 
$\gamma$-ray excesses at 10 GeV or 130 GeV, respectively,
towards the galactic center, as inferred by analyses of Fermi-LAT public 
data \cite{hooper_goodenough,bringmann,weniger,tempel}. They both seem to 
 have a reasonable 
chance to be due to dark matter annihilation. However, the 10 GeV structure may also, like some
previously claimed tentative dark matter signals
 perhaps  be explained by ordinary astrophysical processes. This seems more difficult for a  
$\gamma$-ray line signal, but at present the statistical significance is too weak to claim a discovery. 

The DM explanation of the PAMELA and Fermi-LAT positron data seems to
 require a leptophilic particle of TeV-scale mass and a very much boosted 
cross section. Although this may perhaps be obtained, stretching all 
uncertainties involved \cite{bertone_limits}, and employing Sommerfeld 
enhancement \cite{tracy}, the remaining window seems quite tight. 
If the dark matter particle would indeed be in the TeV energy range, 
indirect detection through $\gamma$-rays in imaging air Cherenkov 
telescopes such as CTA could be a promising method of verification. 

The claimed Fermi-LAT excesses in diffuse emission 
towards the galactic center would of course 
be more plausible if a dark matter signal of mass around 10 GeV, or alternatively at 130 GeV,  were 
established in direct detection experiments. As indicated, the 
situation around 10 GeV at the moment is far from conclusive. The possible line 
signal seems to have, however, a better chance to be promoted to a 
true discovery, if it gains in significance with
 more Fermi-LAT data (which should continue to be accumulated for several
 years). The ``smoking gun'' nature of the 2$\gamma$ signal makes it 
very difficult to find realistic alternative explanations using currently
 known astrophysics. In particular, no acceleration mechanism seems to
 be known that would give such a sharp rise of the flux followed by an equally 
sharp decline at higher energy. It will be very interesting indeed to 
follow how this indication of a 2$\gamma$ line (as first proposed 
in \cite{lbe_snellman} almost 25 years ago) will evolve with time. As we have mentioned, if it
is  confirmed, there are other related processes to search for, and hopefully
also direct detection and LHC searches will give complementary information.

A somewhat different, also complementary, method is indirect detection 
of neutrinos \cite{bertone_indirect}. As the most of the elements in the interior of the Earth
 have spin zero, capture of WIMPS in the Earth takes places through the
 same type of spin-independent scattering that is used bin the CDMS and
 XENON100 detectors. Therefore, the neutrino limits from the interior of 
the Earth are not competitive at the moment. However, the Sun consists 
mostly of hydrogen which means that spin-dependent scattering on protons 
will be important for the capture rate, and consequently for the annihilation
 rate, from the Sun. The spin-dependent limits thus obtained are in many 
cases superior to present-day direct detection limits on spin-dependent 
scattering \cite{icecube}, especially with the low-threshold inset Deep-Core 
and the full 80-string outer IceCube detector now in place. For very high 
mass dark matter particles, the signal from the Sun will however decrease
 due to absorption of the neutrinos in the solar interior.

The times for dark matter research are interesting indeed. Processes that were 
proposed several decades ago will finally  be tested thanks to the rapid 
experimental and observational progress that is currently taking place.
The fact that many experiments now have the reach to probe the parameter
space of several plausible dark matter candidates, will of course also mean
that there will be many false alarms. It is not impossible that all of the
proposed dark matter signals discussed here will turn out to have other explanations. On the other hand, one or more may turn out to survive -- an exciting prospect, indeed.

\section*{Acknowledgements}
I wish to thank many colleagues, especially Felix Aharonian, Gianfranco Bertone, Alexey Boyarsky, Torsten Bringmann, Jan Conrad, Joakim Edsj\"o and Christian Farnier, for many useful discussions. This research was carried out under Swedish Science Research Council (VR) contract no. 621-2009-3915.

\end{document}